\documentclass[twocolumn]{aastex631}
\newcommand{\oiii}{O~III] $\lambda 1663$}

\newcommand{\niv}{N~IV] $\lambda 1488$}

\newcommand{\ergse}{erg~s$^{-1}$}
\newcommand{\ergcse}{erg~cm$^{-2}$s$^{-1}$}
\newcommand{\pyneb}{\texttt{PyNeb}}

\newcommand{\jwst}{JWST}
\newcommand{\lya}{\ifmmode {\rm Ly}\alpha \else Ly$\alpha$\fi}

\def\kmsmpc{km s$^{-1}$ Mpc$^{-1}$}

\def\msun{M_{\odot}}

\shorttitle{The dual nature of GHZ9 at z=10.145}
\shortauthors{Napolitano et al.}
\submitjournal{ApJ}
\usepackage[autopunct=true]{csquotes}
\usepackage{amsmath}
\usepackage{graphicx}
\usepackage{natbib}
\usepackage{scalerel}
\usepackage{makecell}
\usepackage{multirow}
\usepackage{txfonts}
\usepackage{hyperref}  
\usepackage[autopunct=true]{csquotes}
\usepackage{color}
\definecolor{blue}{rgb}{0., 0., 1}
\usepackage{rotating,tabularx}
\hypersetup{colorlinks=true,linkcolor=[rgb]{1.,0.2,0.2},citecolor=[rgb]{0.1,0.4,1.},filecolor=[rgb]{0.7,0.2,0.2},urlcolor=[rgb]{0.7,0.2,0.2}}
\begin{document}

\title{The Dual Nature of GHZ9: Coexisting Active Galactic Nuclei and Star Formation \\Activity in a Remote X-Ray Source at z=10.145}


\correspondingauthor{Lorenzo Napolitano}
\email{lorenzo.napolitano@inaf.it}

\author[0000-0002-8951-4408]{Lorenzo Napolitano}
\affiliation{INAF – Osservatorio Astronomico di Roma, via Frascati 33, 00078, Monteporzio Catone, Italy}
\affiliation{Dipartimento di Fisica, Università di Roma Sapienza, Città Universitaria di Roma - Sapienza, Piazzale Aldo Moro, 2, 00185, Roma, Italy}

\author[0000-0001-9875-8263]{Marco Castellano}
\affiliation{INAF – Osservatorio Astronomico di Roma, via Frascati 33, 00078, Monteporzio Catone, Italy}

\author[0000-0001-8940-6768]{Laura Pentericci}
\affiliation{INAF – Osservatorio Astronomico di Roma, via Frascati 33, 00078, Monteporzio Catone, Italy}

\author[0000-0002-8853-9611]{Cristian Vignali}
\affiliation{Department of Physics and Astronomy (DIFA), University of Bologna, Via Gobetti, 93/2, 40129 Bologna, Italy}
\affiliation{INAF – Osservatorio di Astrofisica e Scienza dello Spazio di Bologna, Via Gobetti, 93/3, 40129 Bologna, Italy}

\author[0000-0001-8121-6177]{Roberto Gilli}
\affiliation{INAF – Osservatorio di Astrofisica e Scienza dello Spazio di Bologna, Via Gobetti, 93/3, 40129 Bologna, Italy}

\author[0000-0003-3820-2823]{Adriano Fontana}
\affiliation{INAF – Osservatorio Astronomico di Roma, via Frascati 33, 00078, Monteporzio Catone, Italy}

\author[0000-0002-9334-8705]{Paola Santini}
\affiliation{INAF – Osservatorio Astronomico di Roma, via Frascati 33, 00078, Monteporzio Catone, Italy}

\author[0000-0002-8460-0390]{Tommaso Treu}
\affiliation{Department of Physics and Astronomy, University of California, Los Angeles, 430 Portola Plaza, Los Angeles, CA 90095, USA}

\author[0000-0003-2536-1614]{Antonello Calabr\`o}
\affiliation{INAF – Osservatorio Astronomico di Roma, via Frascati 33, 00078, Monteporzio Catone, Italy}

\author[0000-0003-1354-4296]{Mario Llerena}
\affiliation{INAF – Osservatorio Astronomico di Roma, via Frascati 33, 00078, Monteporzio Catone, Italy}

\author[0000-0001-9095-2782]{Enrico Piconcelli}
\affiliation{INAF – Osservatorio Astronomico di Roma, via Frascati 33, 00078, Monteporzio Catone, Italy}

\author[0000-0002-4205-6884]{Luca Zappacosta}
\affiliation{INAF – Osservatorio Astronomico di Roma, via Frascati 33, 00078, Monteporzio Catone, Italy}

\author[0000-0002-9572-7813]{Sara Mascia}
\affiliation{INAF – Osservatorio Astronomico di Roma, via Frascati 33, 00078, Monteporzio Catone, Italy}
\affiliation{Dipartimento di Fisica, Università di Roma Tor Vergata, Via della Ricerca Scientifica, 1, 00133, Roma, Italy}

\author[0000-0002-9909-3491]{Roberta Tripodi}
\affiliation{INAF – Osservatorio Astronomico di Roma, via Frascati 33, 00078, Monteporzio Catone, Italy}
\affiliation{IFPU - Institute for Fundamental Physics of the Universe, via Beirut 2, I-34151 Trieste, Italy}

\author[0000-0002-7959-8783]{Pablo Arrabal Haro}
\affiliation{Astrophysics Science Division, NASA Goddard Space Flight Center, 8800 Greenbelt Rd, Greenbelt, MD 20771, USA}

\author[0000-0003-1383-9414]{Pietro Bergamini}
\affiliation{Dipartimento di Fisica, Università degli Studi di Milano, Via Celoria 16, I-20133 Milano, Italy}
\affiliation{INAF – Osservatorio di Astrofisica e Scienza dello Spazio di Bologna, Via Gobetti, 93/3, 40129 Bologna, Italy}

\author[0000-0002-5268-2221]{Tom J.L.C. Bakx} 
\affil{Department of Space, Earth, \& Environment, Chalmers University of Technology, Chalmersplatsen 4 412 96 Gothenburg, Sweden}

\author[0000-0001-5414-5131]{Mark Dickinson}
\affiliation{NSF’s NOIRLab, Tucson, AZ 85719, USA}

\author[0000-0002-3254-9044]{Karl Glazebrook}\affiliation{Centre for Astrophysics and Supercomputing, Swinburne University of Technology, PO Box 218, Hawthorn, VIC 3122, Australia}

\author[0000-0002-6586-4446]{Alaina Henry}
\affiliation{Space Telescope Science Institute, 3700 San Martin Drive, Baltimore, MD 21218, USA}

\author[0000-0003-4570-3159]{Nicha Leethochawalit}
\affiliation{National Astronomical Research Institute of Thailand (NARIT), Mae Rim, Chiang Mai, 50180, Thailand}

\author[0009-0005-7383-6655]{Giovanni Mazzolari}
\affiliation{Department of Physics and Astronomy (DIFA), University of Bologna, Via Gobetti, 93/2, 40129 Bologna, Italy}
\affiliation{INAF – Osservatorio di Astrofisica e Scienza dello Spazio di Bologna, Via Gobetti, 93/3, 40129 Bologna, Italy}

\author[0000-0001-6870-8900]{Emiliano Merlin}
\affiliation{INAF – Osservatorio Astronomico di Roma, via Frascati 33, 00078, Monteporzio Catone, Italy}

\author[0000-0002-8512-1404]{Takahiro Morishita}
\affiliation{IPAC, California Institute of Technology, MC 314-6, 1200 E. California Boulevard, Pasadena, CA 91125, USA}

\author[0000-0003-2804-0648 ]{Themiya Nanayakkara}
\affiliation{Centre for Astrophysics and Supercomputing, Swinburne University of Technology, PO Box 218, Hawthorn, VIC 3122, Australia}

\author[0000-0002-7409-8114]{Diego Paris}
\affiliation{INAF – Osservatorio Astronomico di Roma, via Frascati 33, 00078, Monteporzio Catone, Italy}

\author[0000-0002-2734-7835]{Simonetta Puccetti}
\affiliation{Agenzia Spaziale Italiana-Unita’ di Ricerca Scientifica, Via del Politecnico, 00133 Roma, Italy}

\author[0000-0002-4140-1367]{Guido Roberts-Borsani}
\affiliation{Department of Astronomy, University of Geneva, Chemin Pegasi 51, 1290 Versoix, Switzerland}

\author[0000-0003-2349-9310]{Sofia Rojas Ruiz}
\affiliation{Department of Physics and Astronomy, University of California, Los Angeles, 430 Portola Plaza, Los Angeles, CA 90095, USA}

\author[0000-0002-6813-0632]{Piero Rosati}
\affiliation{INAF – Osservatorio di Astrofisica e Scienza dello Spazio di Bologna, Via Gobetti, 93/3, 40129 Bologna, Italy}
\affiliation{Dipartimento di Fisica e Scienze della Terra, Università degli Studi di Ferrara, Via Saragat 1, I-44122 Ferrara, Italy}

\author[0000-0002-5057-135X]{Eros Vanzella}
\affiliation{INAF – Osservatorio di Astrofisica e Scienza dello Spazio di Bologna, Via Gobetti, 93/3, 40129 Bologna, Italy}

\author[0000-0003-0680-9305]{Fabio Vito}
\affiliation{Scuola Normale Superiore, Piazza dei Cavalieri 7, I-56126 Pisa, Italy}
\affiliation{INAF – Osservatorio di Astrofisica e Scienza dello Spazio di Bologna, Via Gobetti, 93/3, 40129 Bologna, Italy}

\author[0000-0003-0980-1499]{Benedetta Vulcani}
\affiliation{INAF - Osservatorio astronomico di Padova, Vicolo Osservatorio 5, 35122, Padova, Italy} 

\author[0000-0002-9373-3865]{Xin Wang}
\affiliation{School of Astronomy and Space Science, University of Chinese Academy of Sciences (UCAS), Beijing 100049, China}
\affiliation{National Astronomical Observatories, Chinese Academy of Sciences, Beijing 100101, China}
\affiliation{Institute for Frontiers in Astronomy and Astrophysics, Beijing Normal University, Beijing 102206, China}

\author[0000-0001-9163-0064]{Ilsang Yoon}
\affiliation{National Radio Astronomy Observatory, 520 Edgemont Road, Charlottesville, VA 22903, USA}

\author[0000-0002-7051-1100]{Jorge A. Zavala}
\affiliation{National Astronomical Observatory of Japan, 2-21-1, Osawa, Mitaka, Tokyo, Japan}

\begin{abstract}
We present James Webb Space Telescope (\jwst)/NIRSpec PRISM spectroscopic characterization of GHZ9 at z= 10.145 $\pm$ 0.010, currently the most distant source detected by the Chandra X-ray Observatory. The spectrum reveals several UV high-ionization lines, including C~II, Si~IV, N~IV], C~IV, He~II, O~III], N~III], and C~III]. The prominent rest-frame equivalent widths (EW(C~IV)$\simeq$65\AA, EW(O~III])$\simeq$28\AA, EW(C~III])$\simeq$48\AA) show the presence of a hard active galactic nucleus (AGN) radiation field, while line ratio diagnostics are consistent with either AGN or star formation as the dominant ionizing source. GHZ9 is nitrogen-enriched (6--9.5 (N/O)$_{\odot}$), carbon-poor (0.2--0.65 (C/O)$_{\odot}$), metal-poor (Z = 0.01--0.1 Z$_{\odot}$), and compact ($<$ 106 pc), similarly to GN-z11, GHZ2, and recently discovered N-enhanced high redshift objects. We exploited the newly available \jwst/NIRSpec and NIRCam data set to perform an independent analysis of the Chandra data confirming that GHZ9 is the most likely JWST source associated with X-ray emission at 0.5-7 keV. Assuming a spectral index $\Gamma$ = 2.3 (1.8), we estimate a black hole (BH) mass of 1.60 $\pm$ 0.31 (0.48 $\pm$ 0.09) $\times$ 10$^8$M$_{\odot}$, which is consistent either with Eddington-accretion onto heavy ($\geq$ 10$^6$ M$_{\odot}$) BH seeds formed at z=18, or super-Eddington accretion onto a light seed of $\sim$ 10$^2-10^4$ M$_{\odot}$ at z = 25. The corresponding BH-to-stellar mass ratio M$_{BH}$/M$_{star}$= 0.33$\pm$0.22 (0.10$\pm$0.07), with a stringent limit $>$0.02, implies an accelerated growth of the BH mass with respect to the stellar mass. GHZ9 is the ideal target to constrain the early phases of AGN-galaxy coevolution with future multi-frequency observations.

\end{abstract}

\keywords{High-redshift galaxies  --- Primordial galaxies --- Active galactic nuclei --- X-ray active galactic nuclei}
\section{Introduction} \label{sec:intro}
The James Webb Space Telescope (JWST) is revolutionizing our understanding of both galaxies and active galactic nuclei (AGN) in the high-redshift Universe. Several surveys have found a density of bright galaxies at $z>9$, which is significantly larger than  previously predicted \citep[e.g.,][]{Castellano2022b,Castellano2023,Chemerynska2024,Finkelstein2023b,McLeod2024}. In addition, \jwst\ NIRSpec \citep[][]{Jakobsen2022} observations have detected a higher-than-expected number of both broad-line AGN \citep[BLAGN, e.g.,][]{Harikane2023b, Maiolino2023, Matthee2024} and narrow-line AGN \citep[NLAGN, e.g.,][]{Chisholm2024, Curti2024} at high redshift, exceeding the predictions from extrapolated quasar \citep[e.g.,][]{Niida2020, Shen2020} and AGN \citep[e.g.,][]{Finkelstein&Bagley2022} luminosity functions, and from deep X-ray surveys \citep[e.g.,][]{Giallongo2015, Giallongo2019}.
In particular, both \cite{Scholtz2023} and \cite{Mazzolari2024b} showed that NLAGN may represent up to 20\% of the spectroscopically identified galaxy population between 4 $<$ z $<$ 9, based on data from the JADES and CEERS surveys, respectively.\\ 
A comprehensive census of the AGN population at high redshifts is crucial for understanding the origin of the correlation between the physical properties of supermassive black holes (SMBHs) and their host galaxies, as found in the local Universe \citep[e.g.,][]{Kormendy2013, Greene2020}. In addition, a deeper understanding of the demographics and properties of distant AGN is necessary to assess their role in the reionization process \citep[e.g.,][]{Giallongo2015, Madau2024, Dayal2025} and evaluate whether they contribute significantly to the ionizing UV emission from the bright galaxy population at z$\gtrsim$9, as suggested by recent studies \citep[e.g.,][]{Castellano2024, Hegde2024, Maiolino2023, Harikane2024B}. 
However, identifying the AGN population at high redshifts remains challenging.
The demarcation lines between star-forming galaxies (SFGs) and AGN in the classic diagnostic diagrams \citep{Baldwin81, Veilleux87, Kauffmann2003b} are less effective at high redshifts, with the two populations overlapping in the same regions \citep{Kocevski2023, Maiolino2023, Scholtz2023, Ubler2023}. New diagnostic diagrams have been calibrated \citep[e.g.,][]{Calabro2023, Hirschmann2023, Mazzolari2024a} and tested on a few z$>$10 AGN candidates identified by \jwst , such as GN-z11 \citep{Maiolino2024} and GHZ2 \citep{Castellano2024}. \\
The discovery of AGN candidates beyond z=10, when the Universe was less than 450 Myr old, is pushing our exploration closer to the black hole (BH) seeding epoch, offering a unique opportunity to constrain the origin of SMBHs \citep{Natarajan2024, Trinca2024b, Gordon2025, Hu2025, Jeon2025}. 
As recently highlighted by \cite{Taylor2024}, this remote epoch represents a unique window to probe BH seeds, since by z $\sim$ 6 the BH mass function has largely lost memory of its initial seeding phase \citep[see also,][]{Valiante2018}.
The SMBH masses associated with the Chandra X-ray detections of the z$\gtrsim$10 objects UHZ1 \citep{Goulding2023,Bogdan2024}, and GHZ9 \citep{Kovacs2024} have prompted discussions on alternative BH seeding scenarios pathways beyond the standard light and heavy seed scenarios \citep{Inayoshi2020}, including models invoking accretion onto cosmological primordial BH seeds \citep{Dayal2024, Matteri2025, Zhang2025, Ziparo2025}.\\ 
In this paper, we present a detailed analysis of the physical properties of GHZ9, based on the JWST/NIRSpec PRISM data presented in \cite{Napolitano2024B}, which confirmed the object at z=10.145. We examine the contribution of AGN and star formation to the spectrum of this source through rest-frame UV and optical diagnostic diagrams. We also take advantage of the newly available NIRSpec and NIRCam information on the sources in its vicinity to perform an independent analysis of the association with the Chandra X-ray emission, first presented by \citet{Kovacs2024}. GHZ9 provides a unique opportunity to test our understanding of AGN at high redshifts, as it is the most distant X-ray detected AGN known to date and benefits from several rest-frame optical and UV line detections thanks to JWST/NIRSpec.
In this study, we adopt the $\Lambda$CDM concordance cosmological model ($H_0 = 70$ \kmsmpc, $\Omega_M = 0.3$, and $\Omega_{\Lambda} = 0.7$), report all magnitudes in the AB system \citep{Oke1983}, and present equivalent widths (EW) in rest-frame values.
\section{NIRSpec observations and data analysis} \label{sec:Data}
\begin{figure*}
\includegraphics[width=0.81\textwidth]{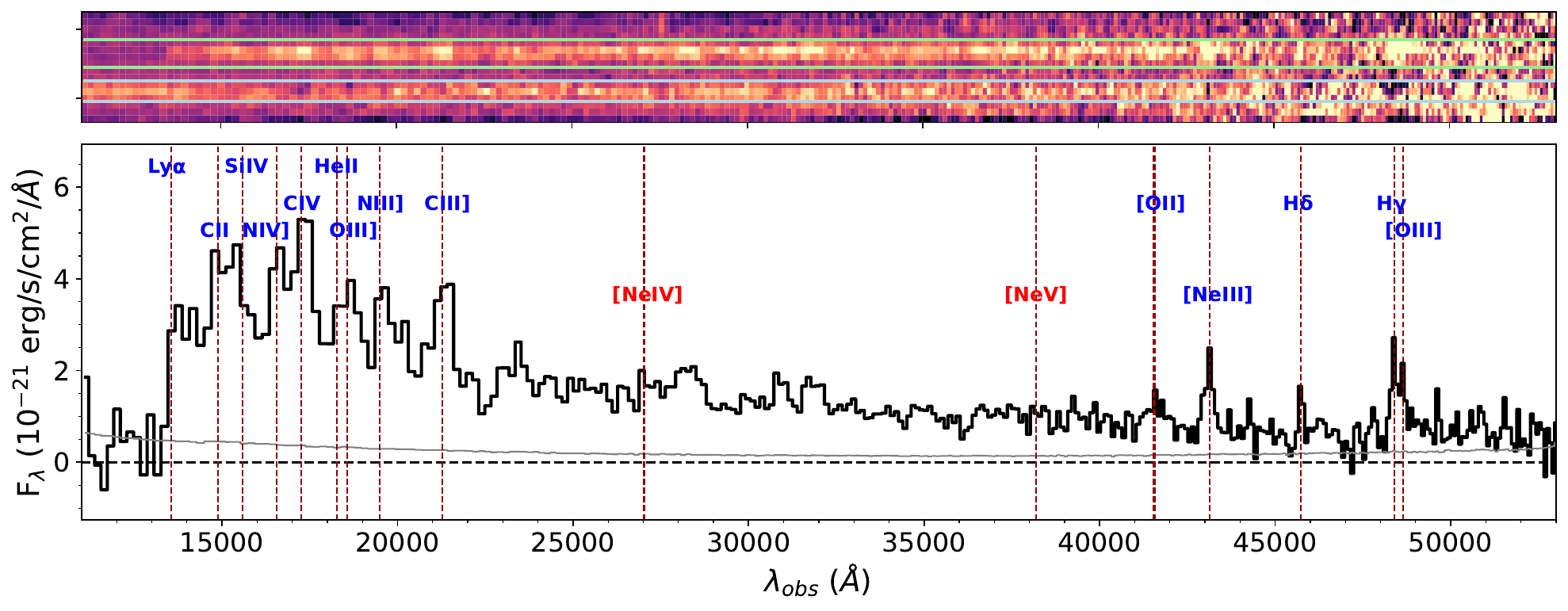}
\includegraphics[width=0.17\textwidth]{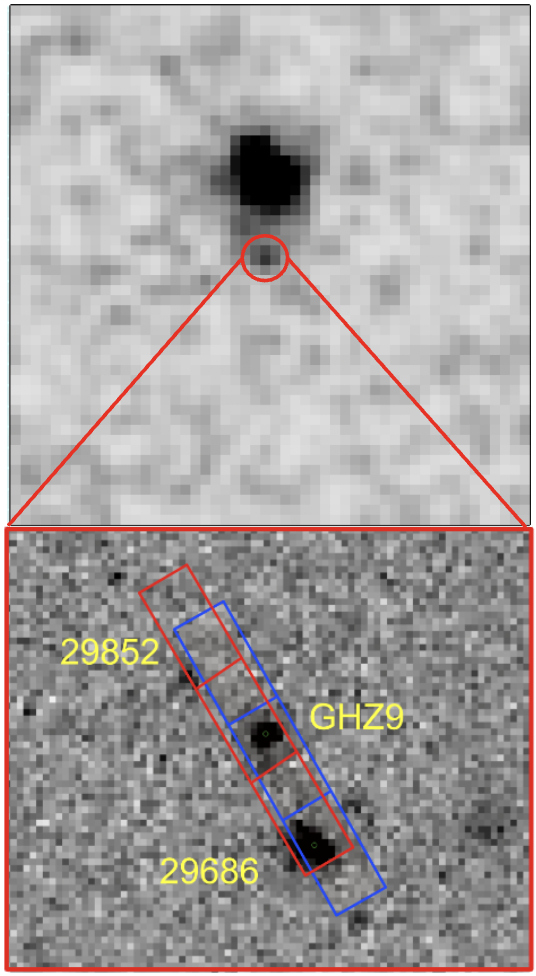}
\caption{Left: Observed 2D (top panel) and 1D spectrum of GHZ9 (bottom panel). The horizontal green and cyan lines enclose the customized extraction regions where we extract the 1D spectra for GHZ9 and ID=29686, respectively. The pipeline error spectrum is reported in gray. Emission lines with an integrated S/N $>$ 3 are marked in blue, while the positions of lines where we have a 3$\sigma$ upper limit are marked in red. The \lya-break feature is shown in blue. Upper Right: 25$\times$25~arcsec$^2$ Chandra image in the 0.5--7~keV band centered on GHZ9, Gaussian smoothed with a 1\arcsec\ FWHM. The source extraction region (radius of 1$''$) is shown in red. Lower right: 2$\times$2~arcsec$^2$ zoom-in F200W image showing the NIRSpec/Prism MSA shutter positions for GHZ9, ID=29686 (z$_{spec}$=1.117), and ID=29852 (z$_{phot}$=0.575), as obtained from the APT tool. The first two visits are shown in blue, and the third visit in red. }
\label{fig:spectra}
\end{figure*}

GHZ9 (R.A. = 3.478756, decl. = -30.345520) was identified as a high-redshift candidate in the GLASS-JWST NIRCam field \citep{TreuGlass2022} by \cite{Castellano2023}. It was observed using NIRSpec in the PRISM-CLEAR configuration as part of the Cycle 2 program GO-3073 (PI: Marco Castellano). The observation utilized three-shutter slits with a three-point nodding pattern for optimal background subtraction, with a total exposure time of 19,701~s over three separate visits.\\
The detailed data reduction, as well as the GHZ9 spectrum and analysis (including redshift determination and line fitting), are presented in \cite{Napolitano2024B}. Briefly, data were processed using the standard calibration pipeline provided by STScI (version 1.13.4) and the Calibration Reference Data System mapping 1197, following the methodology of \cite{Arrabal_Haro2023Nature}, which produces both 2D and 1D flux-calibrated spectra. To correct for potential slit-losses, the NIRSpec spectrum was calibrated against the most recent NIRCam broadband photometry \citep{Merlin2024} by matching the continuum level. Additionally, since the source is magnified by the foreground Abell-2744 cluster, rest-frame quantities for GHZ9 were corrected for magnification \citep[$\mu$ = 1.36,][]{Bergamini2023}. \\ The stellar masses of GHZ9 and of the two foreground interlopers ID = 29686 and ID = 29852 (see Fig.~\ref{fig:spectra}) were estimated using \textsc{zphot} \citep{Fontana2000} as described by \citet[][]{Santini2023}, by fitting the observed HST and JWST photometry\footnote{The photometry of GHZ9 (ID = 29722) and of the two foreground sources is publicly available in the ASTRODEEP catalog \citep{Merlin2024} \url{https://vizier.cds.unistra.fr/viz-bin/VizieR?-source=J/A+A/691/A240}} with \citet[][]{Bruzual2003} templates, assuming delayed star formation histories (SFHs). The contribution from nebular continuum and line emission was included following \citet{Schaerer2009} and \citet{Castellano2014}. 
We measured the half-light radius (r$_e$ = 0.028$''$) in the rest-frame UV using the same procedure adopted by \cite{Mascia2023_CEERS}, with the python software \textsc{Galight}\footnote{\url{ https://github.com/dartoon/galight}} \citep{Ding2020}. Assuming a Sérsic profile, with an axial ratio $q$ between $0.1$ and $1$, and a Sérsic index of $n$ = 1, the fit was performed on the F150W NIRCam image. We visually inspected the result, finding no significant residuals in the luminosity profile.\\
The spectroscopic redshift (z$_{\mathrm{spec}}$ = 10.145 $\pm$ 0.010) was determined from a weighted average of emission line centroids with signal-to-noise ratio (S/N) $>$ 5, calculated via direct integration. We measured the UV slope ($\beta$ = -1.10 $\pm$ 0.12) by fitting a power-law model ($f_{\lambda} \propto \lambda^\beta$) to the continuum flux at 1400--2600 \AA\ rest-frame, after masking any potential emission features within the considered wavelength range. We employed \textsc{emcee} \citep{Foreman_Mackey2013} for Markov chain Monte Carlo (MCMC) analysis.\\
For emission lines with S/N $>$ 3, a Gaussian fit was applied to the continuum-subtracted flux using the \textsc{specutils} package from \textsc{astropy} \citep{Astropy2013} combined with \textsc{emcee}. Unresolved doublets and multiplets were modeled as single Gaussian profiles, while partially blended lines were fitted with double-Gaussian profiles (see Figure A.1 from \cite{Napolitano2024B} for visualization).\\
EW and their uncertainties were calculated based on the integrated flux, continuum flux at the line position, and the spectroscopic redshift. Table~\ref{tab:GHZ9_emission} lists the integrated fluxes and EW of the detected emission lines, along with 3$\sigma$ upper limits through direct integration for undetected features. We report observed line fluxes, the intrinsic values can be obtained by dividing by $\mu$ = 1.36. We note that neither the photometric nor magnification corrections affect the EW or line ratios. \\

\begin{table}[h!]
\centering
\begin{tabular}{lccc}
\hline
\hline
\noalign{\smallskip}
\textbf{Line} & \textbf{Flux} & \textbf{EW}  \\
              & (10$^{-19}$ erg s$^{-1}$ cm$^{-2}$) & (\AA) & \\
\noalign{\smallskip}
\hline
\noalign{\smallskip}
\noalign{\smallskip}
C~II $\lambda \lambda$1335,6  & 8.2 $\pm$ 2.3  & 29 $\pm$ 8   \\  
Si~IV $\lambda \lambda$1394,1403 & 11.4 $\pm$ 2.6  & 41 $\pm$ 9   \\  
\text{N~IV]} $\lambda$1486  & 12.5 $\pm$ 2.0  & 47 $\pm$ 8   \\ 
C~IV $\lambda \lambda$1548,51 & 17.3 $\pm$ 1.9 & 65 $\pm$ 7   \\ 
He~II $\lambda$1640   & 4.5 $\pm$ 2.0  & 18 $\pm$ 8   \\ 
O~III] $\lambda \lambda$1661,66  & 6.9 $\pm$ 1.9  & 28 $\pm$ 8   \\ 
N~III] $\lambda \lambda$1747,49  & 7.9 $\pm$ 1.2  & 33 $\pm$ 5   \\ 
C~III] $\lambda$1908  & 11.0 $\pm$ 1.2  & 48 $\pm$ 5   \\ 
\text{[Ne~IV]} $\lambda$2424 & $<$1.4         & $<$7.8          \\ 
\text{[Ne~V]} $\lambda$3426  & $<$0.78         & $<$7.2          \\ 
\text{[O~II]} $\lambda \lambda$3727,29  & 1.99 $\pm$ 0.39  & 21.2 $\pm$ 4.3  \\ 
\text{[Ne~III]} $\lambda$3869 & 4.11 $\pm$ 0.41 & 47.4 $\pm$ 4.8   \\ 
\text{[Ne~III]} $\lambda$3967 + H$\epsilon$ & $<$1.1 & $<$14   \\ 
H$\delta$ & 1.41 $\pm$ 0.27 & 18.9 $\pm$ 3.7   \\ 
H$\gamma$ & 3.79 $\pm$ 0.49 & 61 $\pm$ 8   \\ 
\text{[O~III]} $\lambda$4363 & 2.9 $\pm$ 0.5 & 46 $\pm$ 9   \\ 
\noalign{\smallskip}
\noalign{\smallskip}
\hline
\hline
\end{tabular}
\caption{Observed flux and rest-frame equivalent width (EW) of detected emission lines for GHZ9. Upper limits are provided at the 3$\sigma$ level. Intrinsic fluxes can be obtained by dividing by the magnification $\mu$=1.36.} \label{tab:GHZ9_emission}
\end{table}

\section{Chandra X-ray observations} \label{sec:X_ray_Data}
 We analyzed all 101 publicly available Chandra X-ray observations of Abell-2744 \citep{Chadayammuri2024}, excluding ObsId=2212 because it was taken with a different CCD (ACIS-S instead of ACIS-I). We reprocessed all of the observations in the whole 0.5--7~keV band where Chandra is sensitive using standard {\sc CIAO} tools and created a mosaic, with the deepest part having an exposure of 2.14~Ms. To ensure accurate astrometry, we matched sources detected in the 0.5–7 keV band to the Gaia DR2 catalog and found that a small translation of 0.27$''$ was required, consistent with \cite{Kovacs2024}. 
GHZ9 has an average off-axis angle of $\sim$6.6$'$ from the deepest part of the X-ray field; the aperture radius including 90\% of the encircled energy fraction (EEF) at 1.5~keV is $\sim$6$''$. The net exposure time, accounting for vignetting, is $\sim$1.76~Ms. 
Given the presence of a nearby ($\sim$4$''$) bright X-ray star, we carefully accounted for its contamination. 

We extracted GHZ9 source counts from a circular region of 1$''$ radius centered on the JWST position in the 0.5--7~keV band image, where a peak of emission is visible in X-rays (see Figure~\ref{fig:spectra}). This region includes only 13\% of the total source counts due to the EEF. To compute the total number of counts, we considered both the nearby star contamination at the location of GHZ9 and the background contribution, evaluated from a nearby source-free circular region of $\sim$100~arcsec$^2$. We also verified that the background is stable over a much larger region. In particular, we found that the background at the position of GHZ9 is consistent with the average background measured at the position of other X-ray sources detected outside the region affected by the diffuse emission from the cluster, whose peak (outskirts) is at $\sim$5.5 (2.8) arcmin from GHZ9. We measured 35.9$\pm$6.9 counts in the 0.5--7~keV band for GHZ9. 

We performed additional checks on the X-ray detection. As a first test, we ran the Chandra detection tool for point-like sources {\sc wavdetect} in a 100$\times$100~arcsec$^2$ region centered on GHZ9, after removing the star contamination. GHZ9 is detected with 23.2$\pm$7.6 counts in the observed-frame 0.5--7 ~keV band (3.3$\sigma$ significance). The flux and luminosity values discussed in the following would be scaled down by $\sim$35\%. As a second check, we evaluated whether the shape of the X-ray continuum of GHZ9 differs from that of the nearby star by computing the ratio between the counts in the soft (S=0.5--2~keV) and hard (H=2--7~keV) bands for both objects. We found this test to be inconclusive due to the large uncertainties resulting in S/H=2.0$^{+1.3}_{-0.8}$ for GHZ9 and 1.2$\pm{0.2}$ for the star. 
As a final check, we divided the Chandra data set into two subsets with comparable exposure time. Unsurprisingly, the source is formally undetected in each individual subset. However, the counts measured within a 1\arcsec\ radius circular region in both subsets are consistent with each other and with the expected signal given the reduced exposure time, thus excluding the possibility that the detection is due to a spurious fluctuation in the counts.

The 0.5--7~keV counts for GHZ9 were then converted into flux and luminosity (both de-magnified) assuming a power-law model with two possible spectral indexes $\Gamma$=1.8 \citep[][]{Piconcelli2005} and $\Gamma$=2.3 \citep[as in][]{Kovacs2024}, which is comparable to the values measured in luminous quasars at z$\sim$6–7.5 \citep{Zappacosta2023}. The assumption of the power-law, related to the main AGN emission process in X-rays (inverse Compton emission of accretion disk photons with energetic electrons in the corona), is required to transform the source count rate, which is a detector-dependent measure, into physical quantities, such as flux and luminosity. In the conversion process, we have considered the Chandra effective area corresponding to Cycle~24, during which most of the Abell~2744 observations were carried out. 
The observed-frame 2--10~keV flux is $\sim$1.3$\times$10$^{-16}$~\ergcse\ if $\Gamma$=2.3 ($\sim$2.0$\times$10$^{-16}$~\ergcse\ if $\Gamma$=1.8). The corresponding rest-frame 2--10~keV luminosity is $\sim$3.8$\times$10$^{44}$~\ergse\ ($\sim$1.8$\times$10$^{44}$~\ergse ). For the observed-frame 0.5--3~keV band, the flux measured, $\sim$2.2$\times$10$^{-16}$~\ergcse, is consistent with the value reported by \cite{Kovacs2024}, assuming the same photon index and magnification factor.



\section{Physical and morphological properties} \label{sec:physical}
The detection of several rest-frame emission lines (Table~\ref{tab:GHZ9_emission}) in GHZ9 provides an opportunity to assess its metallicity (Z), electron temperature (T$_e$), nebular reddening (E(B-V)), nitrogen-to-oxygen (N/O), and carbon-to-oxygen (C/O) abundance ratios. Moreover, from the photometry we assess its morphology and stellar masses. 

Metallicity estimates are derived from 
the Ne3O2 ratio-based calibrations \citep{Shi2007, Maiolino2008, Jones2015, Bian2018, Mingozzi2022, Curti2023}, yielding Z $\in$ [0.01, 0.1] Z$_{\odot}$. This result aligns with recent studies of high-redshift galaxies \citep[e.g.,][]{Goulding2023, Hsiao2023, Carniani2024b, Castellano2024, Maiolino2024, Schouws2024}, which show metal enrichment already at z $>$ 10. The ionization parameter is constrained using the C~IV/C~III] and the C~IV EW relations \citep{Mingozzi2022} and the Ne3O2 relation \citep{Witstok2021}, yielding log\,$U$ = [-1.90, -1.65].\\

\begin{figure}
\includegraphics[width=\linewidth]{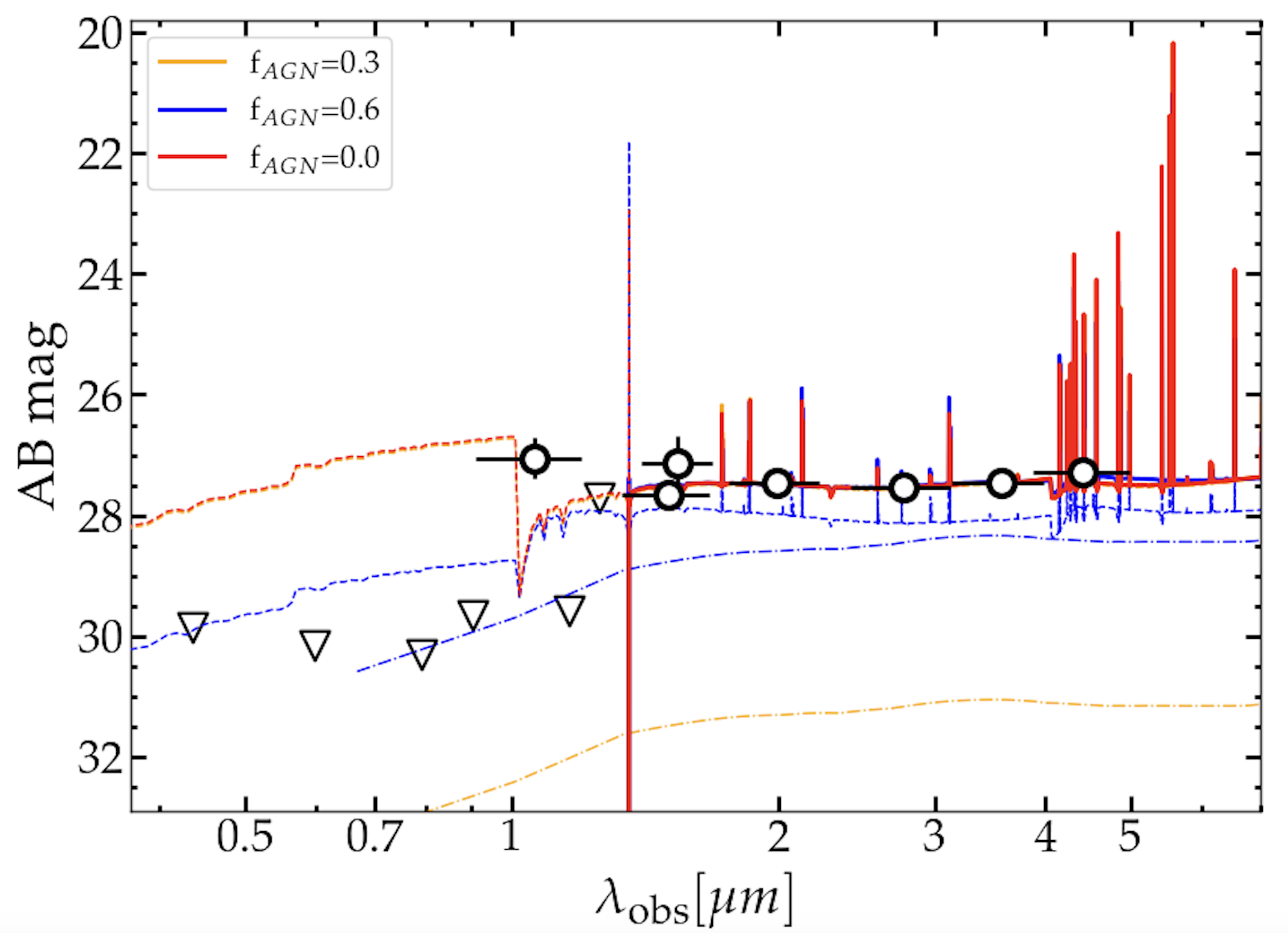}
\caption{Spectral energy distribution of the host galaxy stellar and nebular component (dashed), the AGN component (dot-dashed), and the combined host and AGN fit (solid). Fits are color-coded by increasing AGN fraction used in the modeling. The combined fits are overplotted due to their degeneracy, as discussed in the main text.  Photometric measurements (circles) and 2$\sigma$ upper limits (triangles) are shown from left to right for the following bands: F435W, F606W, F814W, F090W, F105W, F115W, F125W, F150W, F160W, F200W, F277W, F356W, and F444W, as reported in the ASTRODEEP catalog from \cite{Merlin2024}.}
\label{fig:SEDfit}
\end{figure}

We perform an AGN+SFG SED fit using the \textsc{dale2014} module \citep{Dale2014} from \textsc{CIGALE} \citep{Boquien2019} and considering the available HST and JWST photometry (Figure~\ref{fig:SEDfit}). 
%
We adopt a flexible SFH with two parametric components to take in consideration the presence of both old and young stellar populations. Namely, we include a \textquote{delayed} component of age $\geq$200 Myr and allow for a recent exponential burst with timescale 1$\leq\tau\leq$50 Myr and age between 5 and 100 Myr. We assume a \citet{Chabrier2003} initial-mass function (IMF), a \cite{Calzetti2000} extinction law and we restrict metallicity and ionization parameter to the ranges estimated from the NIRSpec spectrum. Namely, the gas metallicity can be 2\%, 5\%, and 10\% the solar value, while the stellar metallicity is fixed to 2\% solar, and log\,$U$ can vary in the range [-1.90, -1.60] with 0.1 steps. 
The AGN component is parameterized by the AGN fraction (f$_{AGN}$), defined as the ratio of the AGN luminosity to the total AGN and dust luminosities. We initially performed the fit with the AGN fraction set as a free parameter and obtained a best-fit likelihood-weighted value of f$_{AGN}$=0.30 (68\% confidence region 0.01--0.6), 
resulting in a lens-corrected stellar mass of 4.9$^{+3.6}_{-3.2}$ $\times$ 10$^8$ M$_{\odot}$. 
To better understand the impact of the AGN on the stellar-mass estimate, we repeated the fit fixing f$_{AGN}$ at discrete values within the 0--0.6 range. The lens-corrected stellar mass was found to vary between 3.3$^{+2.4}_{-2.3}$ 
and 7.2$^{+3.0}_{-3.8}$ 
$\times$ 10$^8$ M$_{\odot}$, for the maximum and null AGN contribution, respectively. The best-fit nebular reddening is E(B–V) = 0.32 $\pm$0.18.

The relatively red UV slope we observe ($\beta$ = -1.10 $\pm$ 0.12) is consistent with either a scenario of significant dust obscuration, corresponding to a nebular E(B-V) = 0.315 $\pm$  0.047 \citep{Castellano2014} or the AGN nature of the source \citep[$\langle \beta \rangle _{AGN}$ = -1.5 $\pm$ 0.7, ][]{Greene2024}. To investigate nebular reddening from emission lines, we examined the Balmer decrement from the observed ratio between H$\gamma$ and H$\delta$. Using the intrinsic H$\gamma$/H$\delta$ ratio of 1.81 from \cite{Osterbrock2006}, assuming case B recombination at a density of $n_e$ = 100 cm$^{-3}$ and a temperature of $T_e$ = 10$^4$ K, and the reddening curve from \cite{Calzetti2000}, we derived E(B-V) = 1.6 $\pm$ 1.0. This value is consistent with a scenario of significant dust obscuration and compatible with the reddening found from the SED fit.

We measured the electron temperature from the flux ratio between O~III] $\lambda \lambda$1661,66 and [O~III] $\lambda$4363 lines, correcting for dust attenuation using the \cite{Calzetti2000} law and the E(B-V) value derived from the SED fitting. Assuming an electron density of 10$^3$ cm$^{-3}$, we used \pyneb\ to compute a conservative 2$\sigma$ (1$\sigma$) lower limit based on the corresponding lower bounds of the flux ratio. This yields T$_e$ = 2 $\times$ 10$^4$ K (4 $\times$ 10$^4$ K), in good agreement with recent results for high-redshift AGN \citep[e.g.,][see their Figure~4]{Tripodi2024}. The large uncertainty in E(B–V), combined with the limited spectral resolution that prevents decomposition of the O~III] $\lambda \lambda$1661,66 components, limits a more precise determination of T$_e$.


In GHZ9, the high observed EW of N~III] (33 $\pm$ 5 \AA) and N~IV] (47 $\pm$ 8 \AA) suggest a nitrogen-enriched nature, with values exceeding those reported for N-enriched galaxies in recent studies \citep[e.g.,][]{Bunker2023B, Castellano2024, Ji2024, Schaerer2024, Topping2024, Hayes2025}. We therefore further investigated the N-enriched nature of GHZ9, by computing the nitrogen-to-oxygen (N/O) abundance ratio, approximating it as (N$^{2+}$ + N$^{3+}$)/O$^{2+}$ using \pyneb\ \citep{Luridiana2012, Luridiana2015}.
We use the measured line ratios \niv/\oiii\ and N~III] $\lambda 1750$/\oiii\ and consider a range of electron densities ([10$^3$, 5 $\times$ 10$^3$, 10$^4$, 5 $\times$ 10$^4$, 10$^5$, and 5 $\times$ 10$^5$]~cm$^{-3}$) and temperatures ([1.5, 2, 2.5, 3] $\times$ 10$^4$~K). We performed a Monte Carlo analysis by perturbing the observed fluxes by their corresponding uncertainties 1000 times. The resulting N/O values range from -0.08 (for ne= 5 $\times$ 10$^5$cm$^{-3}$ and Te=15,000 K) to 0.12 (for ne=10$^3$cm$^{-3}$ and Te=30,000 K), which are $\sim$ 6--9.5 times higher than the solar N/O value \citep[log(N/O)$_{\odot}$ = -0.86;][]{Asplund2009}. \\
We also estimate the C/O abundance using the C$^{2+}$/O$^{2+}$ ratio, with the same method and density and temperature ranges as for the N/O. We apply the ionization correction factor (ICF) from \cite{Berg2019}, which depends on gas metallicity (Z) and ionization parameter (log\,$U$).
For the measured ranges of $Z$ and log\,$U$, the resulting ICF ranges from 1.10 to 1.32. Using these parameters, we derive log(C/O) values ranging from -0.96 (for n$_e$= 5 $\times$ 10$^5$cm$^{-3}$ and Te=15,000 K) to -0.45 (for n$_e$=10$^3$cm$^{-3}$ and Te=30,000 K), which are $\sim$ 0.2--0.65 times the solar C/O value \citep[log(C/O)$_{\odot}$ = -0.26;][]{Asplund2009}. 
We note that the exceptionally high C~III] EW = (48 $\pm$ 5) \AA\ in GHZ9 is matched by only one other source in the literature, UNCOVER-45924, a BLAGN at z=4.5 \citep{Greene2024, Treiber2024}. UNCOVER-45924 is associated with a secure [Ne~V] detection, whereas GHZ9 lacks high-ionization lines ($>$ 60~eV) such as [Ne~IV] and [Ne~V], for which we provide a 3$\sigma$ upper limit in Table~\ref{tab:GHZ9_emission}. However, we find significant (S/N $>$ 3) detections of the C~II and Si~IV multiplets. As discussed in \cite{Maiolino2024} for GN-z11, these lines are commonly observed in AGN spectra \citep[e.g.][]{Wu2022}. Higher resolution observations are needed to better constrain the multiplets and determine if we can detect a broad component from permitted lines.

In terms of morphology, GHZ9 is compact with a half-light radius r$_e$ = 0.028$''$, corresponding to 99 pc after correcting for the lensing effect. Given the instrument's spatial resolution of 0.03$''$ we adopt a conservative upper limit of 106 pc. We note that GHZ9 would be classified among compact galaxies with strong high-ionization lines \citep[the \textquote{strong N~IV \& compact} cloud of galaxies at z$>$9, as defined in][]{Harikane2024B}, further indicating that potential AGN activity affects the observed high density of bright galaxies at these high redshifts \citep[][]{Castellano2023, Napolitano2024B}.

\section{Evidence of AGN emission from UV and optical line diagnostics} \label{sec:AGNvsSF}
The detection of several rest-frame emission lines in GHZ9 (Table~\ref{tab:GHZ9_emission}) provides an opportunity to assess whether its primary ionizing source is an AGN or stellar populations. To achieve this, we use a combination of UV emission line ratios and EW diagnostics from the literature \citep[][]{Feltre2016, Gutkin2016, Nakajima2018a, Hirschmann2019}. Specifically, we analyze the EW of key UV metal lines (C~IV $\lambda \lambda$1548,51 , O~III] $\lambda \lambda$1661,66 , N~III] $\lambda \lambda$1747,49 , C~III] $\lambda$1908) and their ratios to the He~II $\lambda$1640 recombination line because they are sensitive indicators to the hardness of the ionizing radiation. \\
We compare these measurements to the photoionization models by \cite{Nakajima2022} (hereafter NM22), constructed with the \textsc{CLOUDY} code \citep[][]{Ferland2013}. These models consider a variety of ionization sources, including SFG, AGN, Population III stars, and direct collapse black holes and span a broad range of physical parameters, including gas metallicities, ionization parameters, and gas densities \citep[see also][]{Nakajima2018a}. They also incorporate BPASS stellar population synthesis models \citep{Eldridge2017}. In our analysis, we restrict the comparison to the NM22 AGN and SFG grids that are consistent with the physical properties inferred for GHZ9. In particular, we explore models with -2 $\leq$ log\,$U$ $\leq$ -1.5, 10$^{-4}$$\leq$ Z$_{gas}$ $\leq$ 2 $\times$ 10$^{-3}$, and n$_H$=10$^3$cm$^{-3}$. The models account for varying C/O and N/O abundance ratios as a function of metallicity \citep{Dopita2006, Nakajima2018a}. \\
We apply the same parameter constraints in log\,$U$ and Z$_{gas}$ when comparing GHZ9 to the photoionization models by \cite{Feltre2016} (hereafter F16) and \cite{Gutkin2016} (hereafter G16). In this case, the SFG models are further restricted to 0.14 $\leq$ (C/O)/(C/O)$_{\odot}$ $\leq$ 0.72. The nitrogen abundance follows the prescription of \cite{Groves2004} in their grid. To aid interpretation, we include the UV diagnostic demarcation lines from \cite{Hirschmann2019} \citep[see also,][]{Hirschmann2023}, which distinguish between AGN-dominated, SFG-dominated, and composite sources. These diagnostics were designed to maximize AGN classification purity (approximately 90\%) up to z=8, which corresponds to 180 Myr of galaxy evolution from GHZ9. 
Additionally, we employ the optical diagnostic diagram introduced by \cite{Mazzolari2024a}, based on [O~III] $\lambda$4363 / H$\gamma$ versus [Ne~III] $\lambda$3869 / [O~II] $\lambda \lambda$3727,29. While [Ne~III]/[O~II] (Ne3O2) traces the ionization state of the interstellar medium, [O~III] $\lambda$4363 provides information about the electron temperature, thereby offering insight into the energy output of the ionizing source. We note that both the \cite{Hirschmann2019} and \cite{Mazzolari2024a} diagnostics were derived from a broader space  than the one strictly matching GHZ9’s properties \citep[see Table 1 in][]{Hirschmann2017}. 
Figure~\ref{fig:AGNvsSF_diagrams} shows the position of GHZ9 in these diagnostic diagrams and compares it with other spectroscopically confirmed AGN candidates at z $>$ 8.5, including UNCOVER-20466 \citep[z=8.50;][]{Kokorev2023}, CEERS-1019 \citep[z=8.68;][]{Larson2023}, GS-z9-0 \citep[z=9.43;][]{Curti2024}, GN-z11 \citep[z=10.6;][]{Bunker2023B, Maiolino2024}, and GHZ2 \citep[z=12.34;][]{Castellano2024}. For comparison we show the AGN and SFG photoionization models from NM22, F16, and G16.

As discussed by \cite{Castellano2024} and \cite{Maiolino2024}, the interpretation of whether the observed high-redshift AGN candidates, including GHZ9, are primarily AGN- or SFG-dominated is strongly model-dependent. Here, we detail our analysis based on the diagnostic diagrams presented in Figure~\ref{fig:AGNvsSF_diagrams}. In the C~III]/He~II versus C~IV/C~III] and the C~III]/He~II versus O~III]/He~II diagnostics, GHZ9 lies in the composite region, consistent within 1$\sigma$ with both the explored AGN and SFG models predictions. Therefore, these diagnostics do not allow us to determine the dominant ionizing source powering the observed emission lines. \\
In the optical [O~III]/H$\gamma$ versus [Ne~III]/[O~II] diagram, GHZ9 lies in the AGN-dominated region, though it is consistent within 2$\sigma$ with the composite region and with both AGN and SFG predictions by F16 and G16. When using the NM22 models, however, only the AGN scenario is compatible with the observed emission lines and their uncertainties.\\
The AGN-dominated nature of GHZ9 is most strongly supported by the EW based diagnostics of C~IV, O~III], and C~III], which are only provided by NM22. Although GHZ9 is formally consistent within 1$\sigma$ with both the AGN-dominated and composite regions, within their uncertainty the high EW observed are only reproduced by AGN models with physical parameters comparable to those of GHZ9.\\
Finally, using the nitrogen-based diagnostics (C~III]/He~II versus N~III]/He~II and EW NIII] versus N~III]/He~II) the position of GHZ9 is compatible with SFG-dominated and composite regions. However the observed EW of N~III] exceeds theoretical predictions by more than an order of magnitude. No AGN or SFG model explored in our analysis reproduces such extreme EW values. This suggests that these diagrams primarily reflect the N-enriched nature of GHZ9 rather than constraining its ionizing source. To our knowledge, no existing models are designed to distinguish AGN from star formation emission in N-enhanced sources.

\begin{figure*}[ht!]
    \centering
    \begin{minipage}{0.49\textwidth}
        \includegraphics[width=\linewidth]{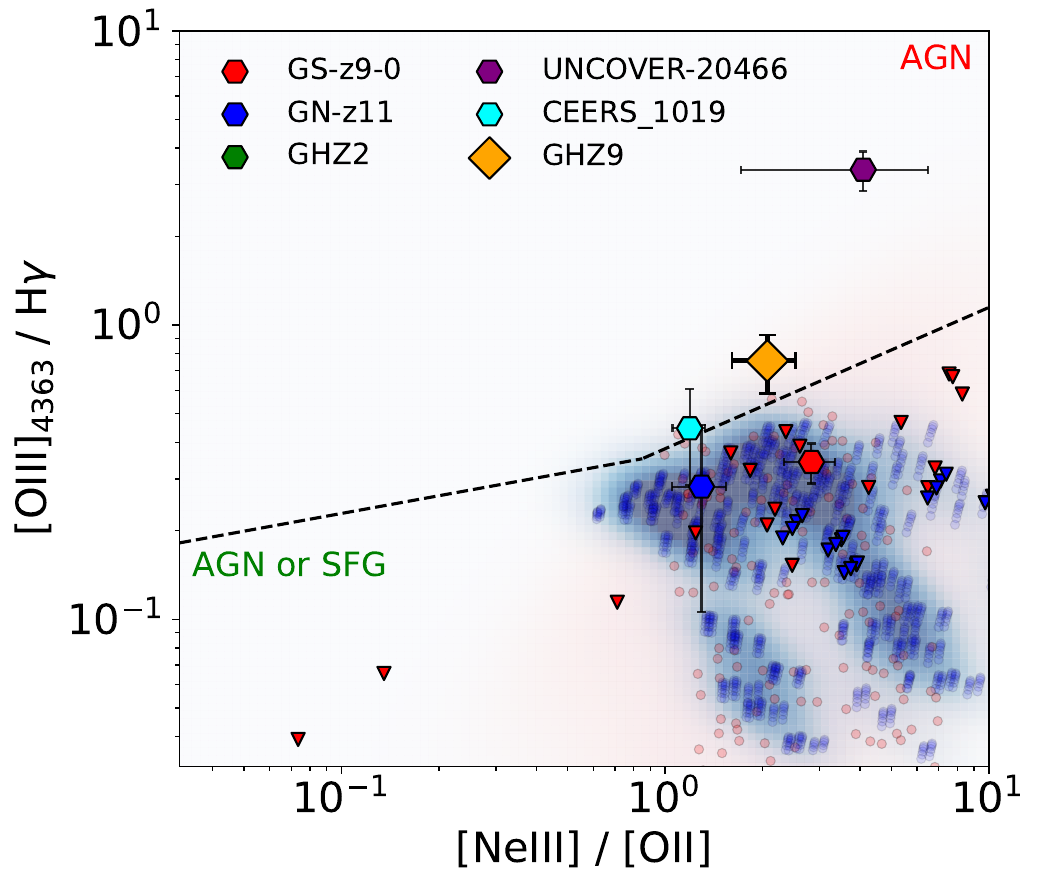}
    \end{minipage}
    \begin{minipage}{0.49\textwidth}
        \includegraphics[width=\linewidth]{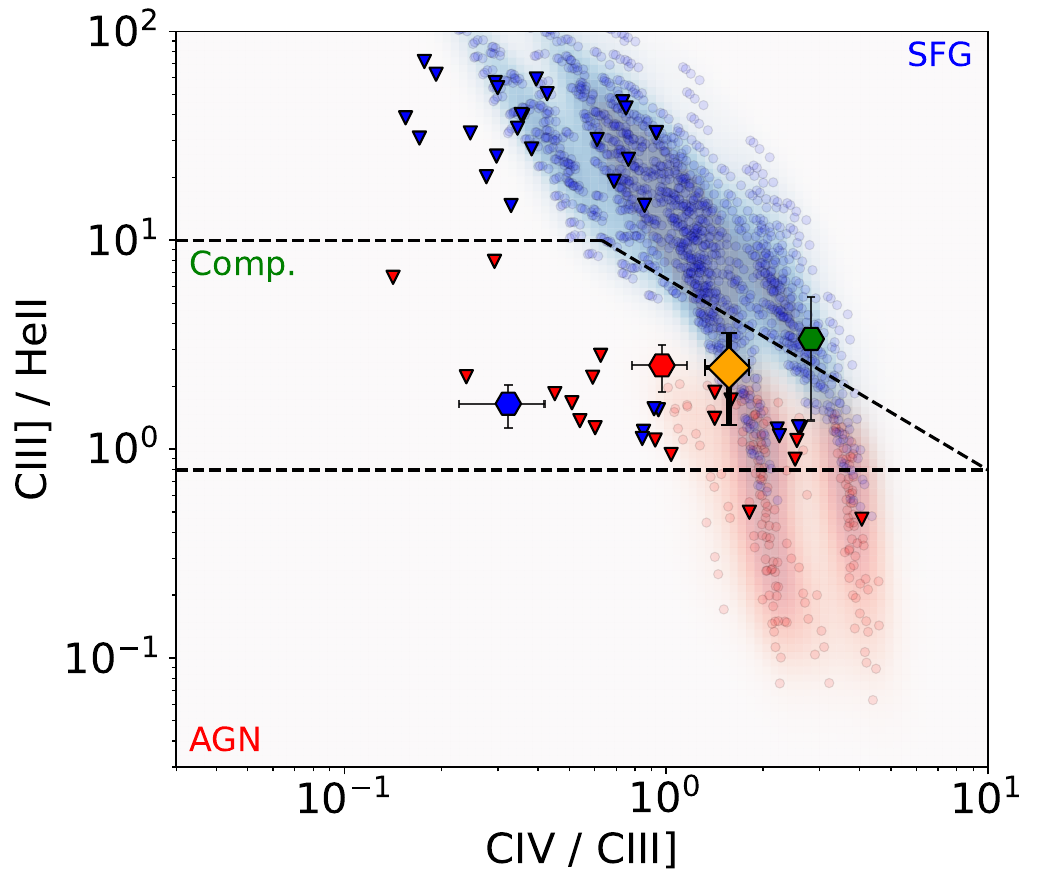}
    \end{minipage}
    
    \medskip
    
    \begin{minipage}{0.49\textwidth}
        \includegraphics[width=\linewidth]{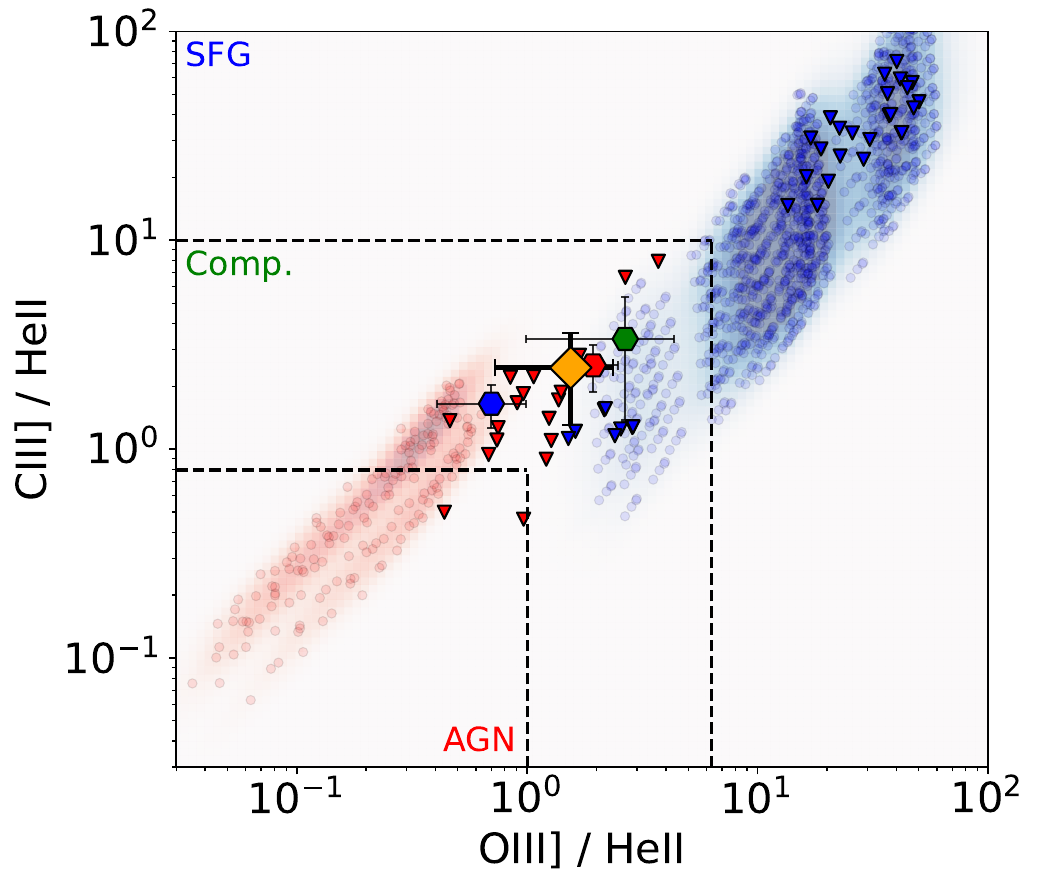}
    \end{minipage}
    \begin{minipage}{0.49\textwidth}
        \includegraphics[width=\linewidth]{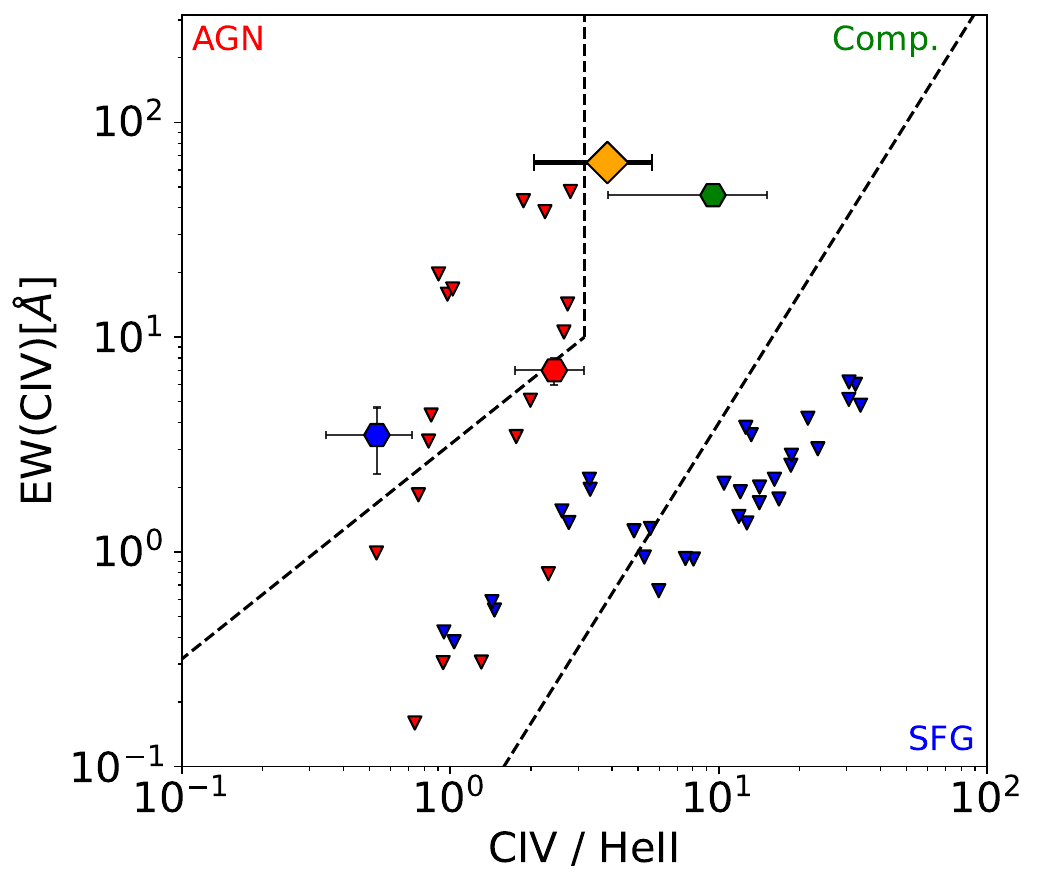}
    \end{minipage}
    
    \medskip
    
    \begin{minipage}{0.49\textwidth}
        \includegraphics[width=\linewidth]{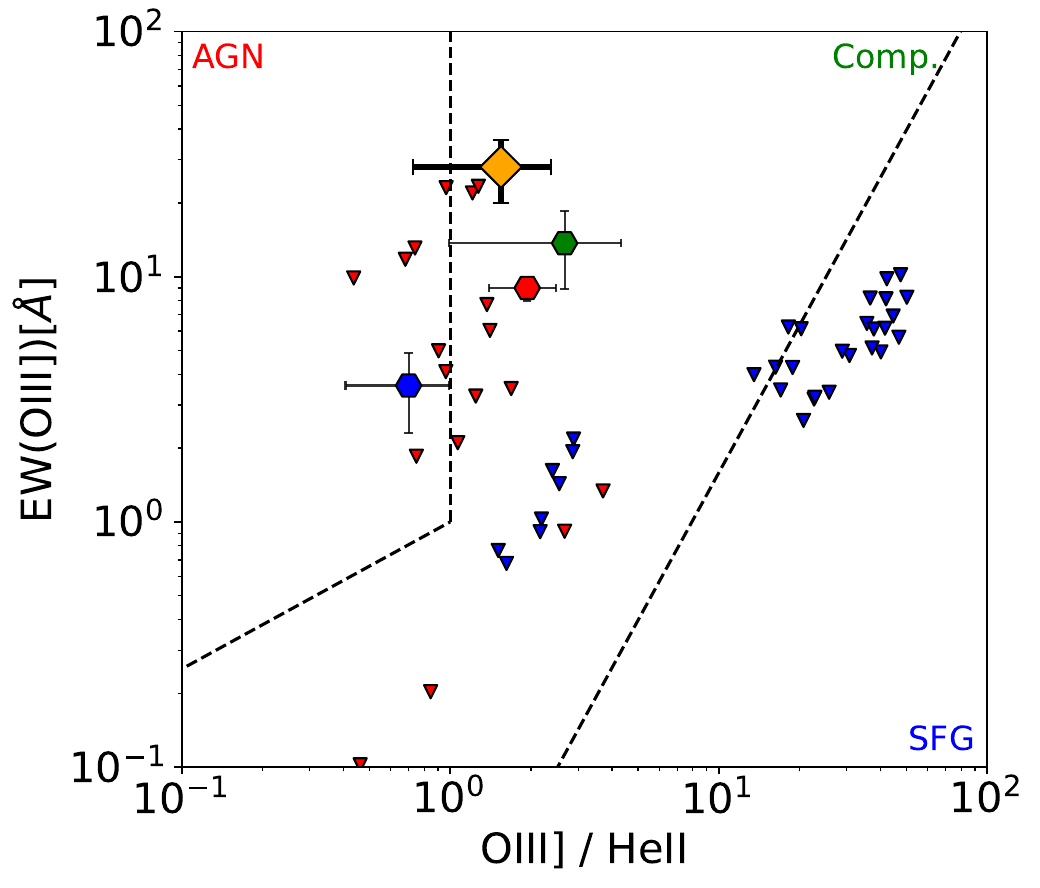}
    \end{minipage}
    \begin{minipage}{0.49\textwidth}
        \includegraphics[width=\linewidth]{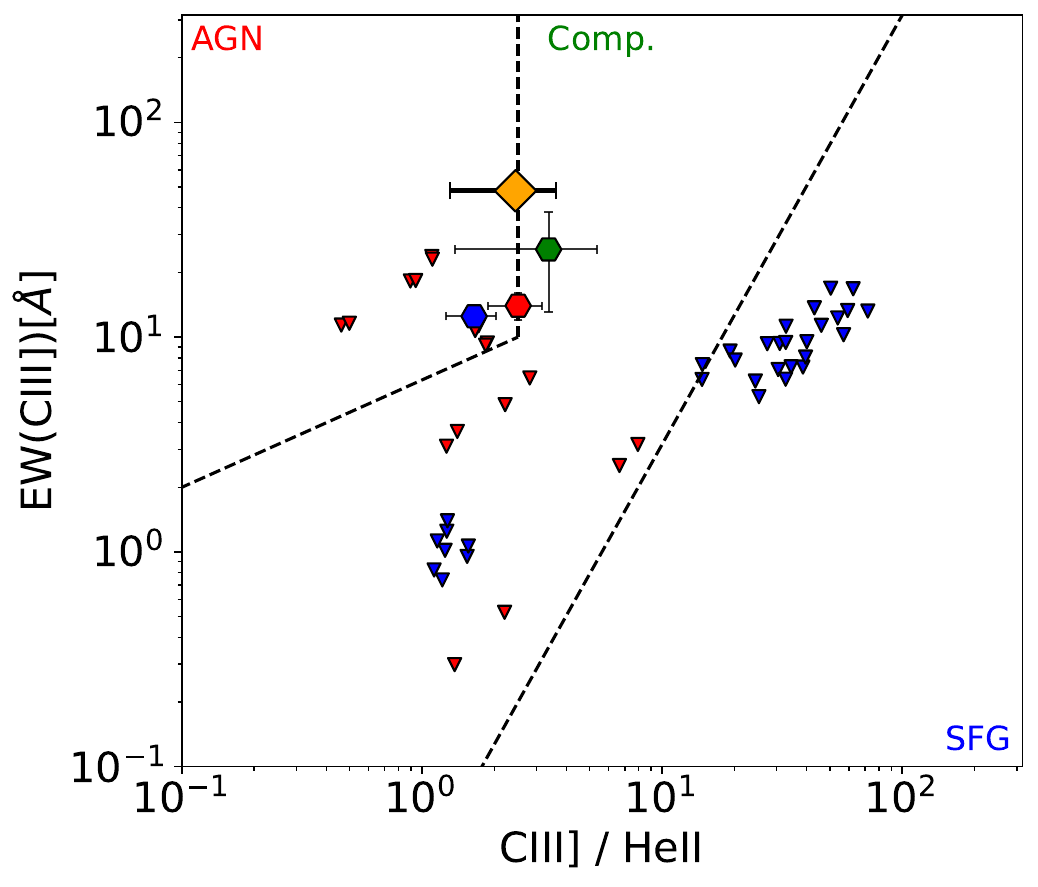}
    \end{minipage}
    
\end{figure*}

\begin{figure*}[]
    \begin{minipage}{0.49\textwidth}
        \includegraphics[width=\linewidth]{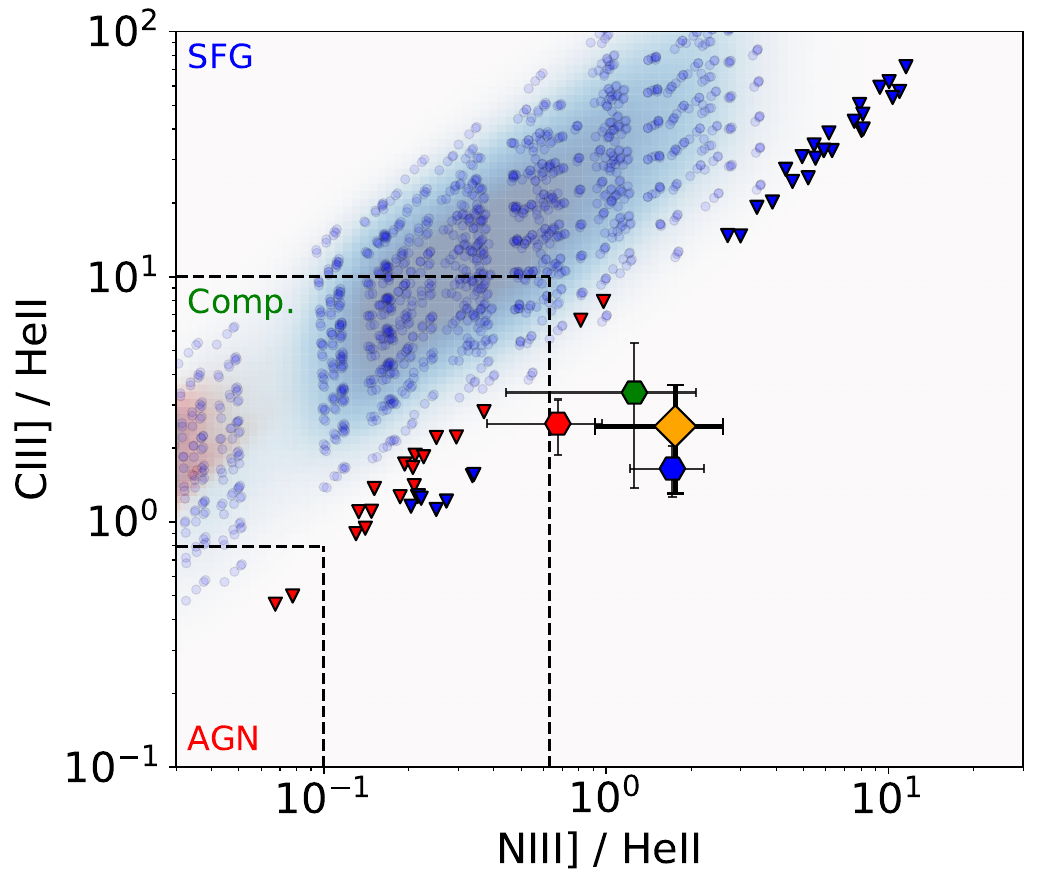}
    \end{minipage}
    \begin{minipage}{0.49\textwidth}
        \includegraphics[width=\linewidth]{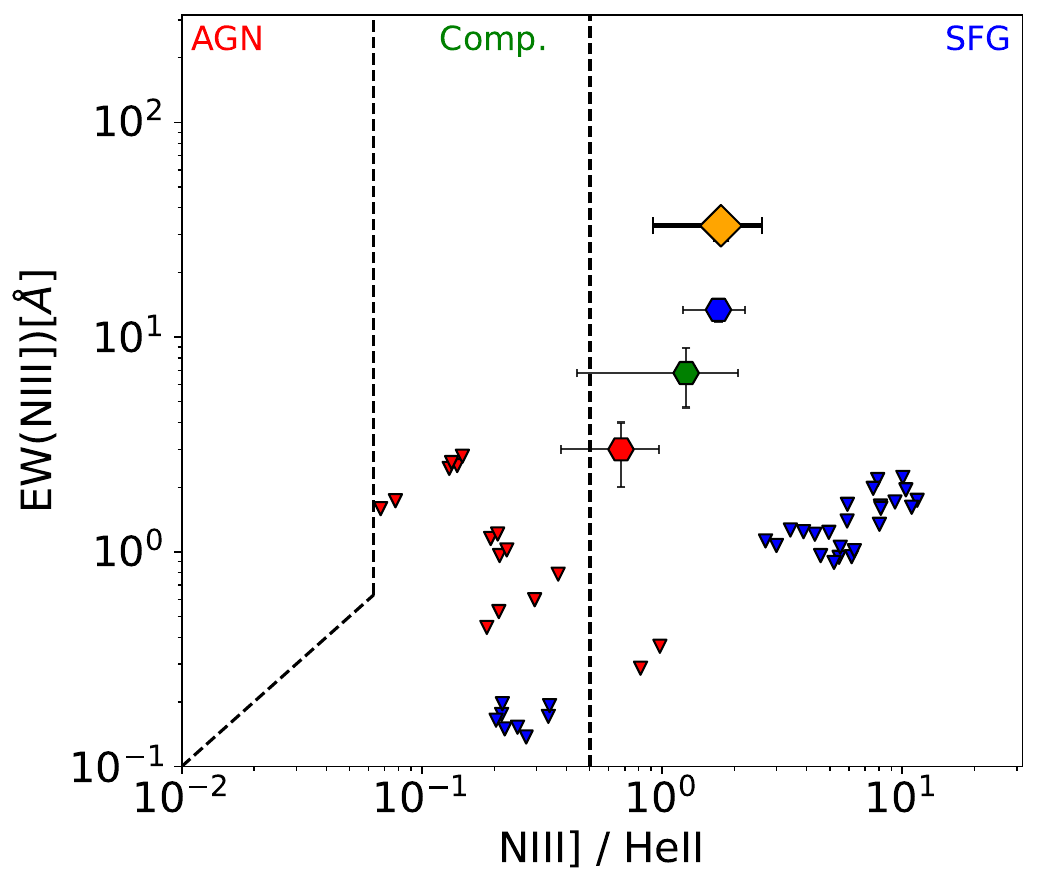}
    \end{minipage}
    \caption{Diagnostic diagrams based on flux ratios and EW. The top left panel shows the legend for each high-redshift AGN candidate included. AGN and SFG models from NM22 are plotted as red and blue triangles, respectively, while F16 and G16 models are shown as circles with shaded backgrounds following the same color scheme. Demarcation lines used to separate AGN-dominated, SFG-dominated, and composite sources follow the criteria from \cite{Hirschmann2023} and \cite{Mazzolari2024a}.}
    \label{fig:AGNvsSF_diagrams}
\end{figure*}

\section{Nature of the X-ray emission} \label{sec:Xray}
We first assessed the reliability of the association between the X-ray detection and GHZ9. As discussed in Section~\ref{sec:X_ray_Data}, the Chandra X-ray counts have been measured within a circular region of 1$''$ radius where two other NIRCam-detected objects are found. The sources are ID=29686 and ID=29852 from the catalog by \citet{Merlin2024}, located $\sim$0.5$''$ from GHZ9. The spectrum of ID=29686 was serendipitously observed in the same slit as GHZ9. Based on the H$\alpha$ emission, the unresolved [OIII]$\lambda \lambda$4959, 5007, and [OII]$\lambda \lambda$3727, 3729 doublets, we determine the spectroscopic redshift to be z$_{\mathrm{spec}}$ = 1.117 $\pm$ 0.006, with a corresponding magnification $\mu$ = 1.23 from \cite{Bergamini2023}. We further characterized this source to investigate, using the log\,([OIII]$\lambda 5007$ / H$\beta$) versus log\,(M/M$_{\odot}$) diagnostic \citep[MEx diagram, ][]{Juneau2011}, whether it could be compatible with an AGN nature. From SED fitting analysis, we find its stellar mass to be log\,(M/M$_{\odot}$) = 7.65$^{+0.08}_{-0.02}$ and a null E(B-V) measurement. The [OIII]$\lambda 5007$ and H$\beta$ fluxes were obtained assuming no dust correction and case B recombination with a density $n_e$ = 100 cm$^{-3}$ and temperature $T_e$ = 10,000 K. The results indicate a star-forming nature, with the AGN scenario disfavored at a significance level of 4.5 $\sigma$.

Instead, ID=29852 is an ultrafaint source ($m_{F200W}=$29.7), whose best-fit model corresponds to a low-mass (log\,(M/M$_{\odot}$) = 6.3$^{+0.6}_{-1.0}$) passive (sSFR$<$10$^{-11}$ yr$^{-1}$) galaxy at z$_{phot}$ = 0.575. Assuming the X-ray emission to be associated with this object, the resulting BH mass would be $\sim$ 2.6 $\times$ 10$^4$ M$_{\odot}$. In this scenario the galaxy would be hosting an intermediate-mass BH (10$^2$--10$^5$ M$_{\odot}$), a rare class of objects for which no direct identification has been obtained beyond the local Universe \citep[e.g.,][]{Greene2020, Boorman2024}. 

We finally assessed whether stellar processes in ID=29686 and ID=29852 could account for the observed X-ray emission by using the SFR values derived from the SED fitting and applying the X-ray luminosity relation for SFGs from Equation~(14) in \cite{Lehmer2016}. We find that the expected X-ray luminosities (10$^{39}$ and 10$^{37}$\ergse , respectively) are more than three orders of magnitude lower than the observed ones (10$^{42}$ and 10$^{41}$\ergse), effectively ruling out a stellar origin for the X-ray emission in both cases. Similarly, for GHZ9, we exclude the possibility that the X-ray emission is due to stellar processes because the measured X-ray luminosity (Section~\ref{sec:X_ray_Data}) for an object with its stellar mass would require a SFR more than two orders of magnitude higher than what is estimated for this source.

Therefore, we conclude that GHZ9, whose spectrum is consistent with the presence of AGN activity, is the most likely source associated with the X-ray Chandra detection.

\begin{figure*}[ht!]
\centering
\includegraphics[width=\linewidth]{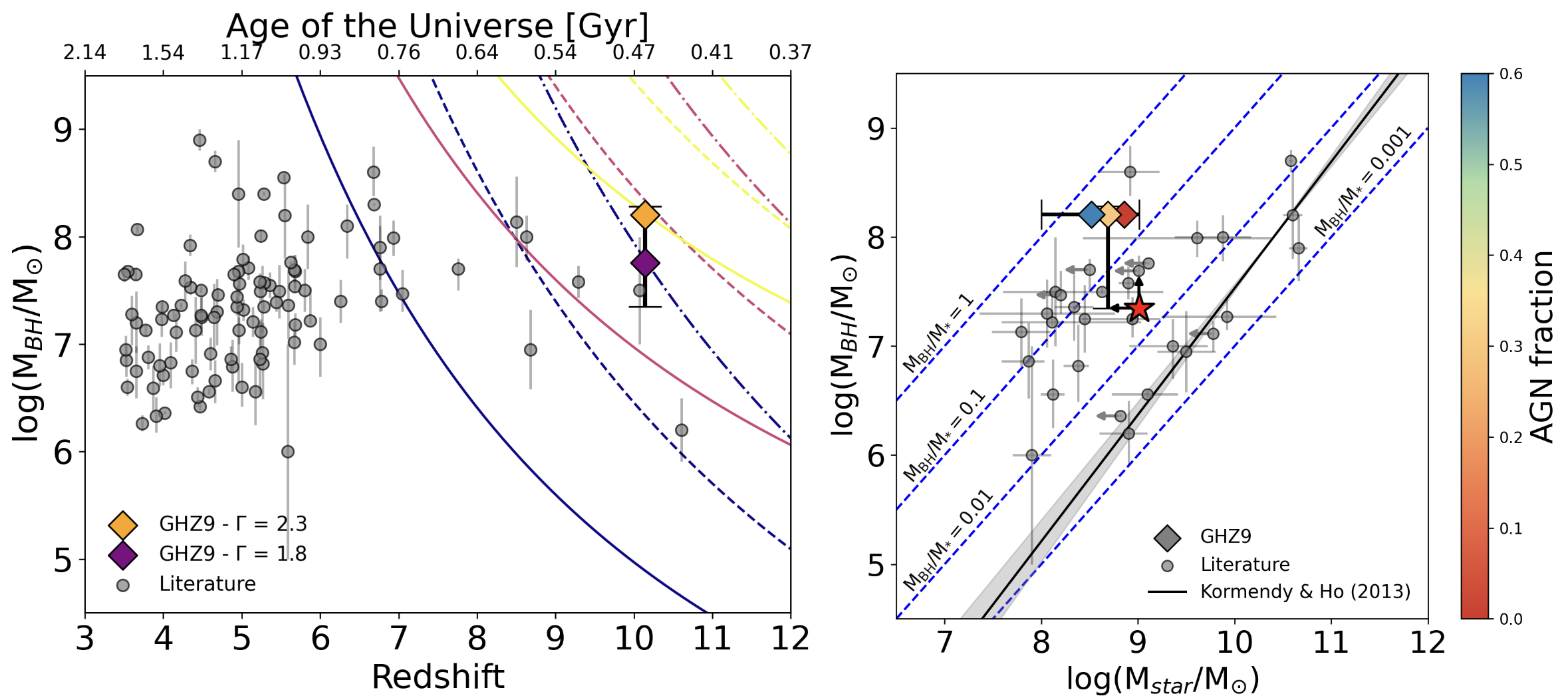}
\caption{Left: BH mass as a function of observed redshift. We report evolutionary models of BH mass that differ based on the initial mass seed and accretion rate. Yellow, pink, and blue colors represent 10$^6$ M$_{\odot}$, 10$^4$ M$_{\odot}$, and 10$^2$ M$_{\odot}$, respectively. Solid, dashed, and dot-dashed lines refers to the 1, 1.5, and 2.0 Eddington accretion rates. We present the inferred BH mass of GHZ9 based on two different spectral indices, as discussed in the main text. Right: BH mass versus stellar mass of the host galaxy. We show the stellar mass of GHZ9 color-coded by increasing AGN fractions used in the SED fitting, while adopting the BH mass solution for a steep spectral index $\Gamma$ = 2.3. The error bars for GHZ9 show the statistical uncertainties for the extreme cases, where the stellar and BH masses are at their minimum and maximum values. The red star shows our stringent lower limit M$_{BH}$/M$_{star}=$0.02, as discussed in the main text.  The M$_{BH}$ -- M$_{bulge}$ relation obtained by \cite{Kormendy2013} is indicated by the solid black line and gray shaded region. The gray symbols show estimates from observed JWST active galaxy at z $>$ 3.5 from the literature: \cite{Carnall2023}, \cite{Goulding2023}, \cite{Harikane2023b}, \cite{Kocevski2023}, \cite{Kokorev2023}, \cite{Larson2023}, \cite{Maiolino2023}, \cite{Ubler2023}, \cite{Chisholm2024}, \cite{Furtak2024}, \cite{Greene2024}, \cite{Juodzbalis2024}, \cite{Matthee2024}, \cite{Taylor2024}, \cite{Tripodi2024}, \cite{Naidu2025A}, and \cite{Taylor2025}.}
\label{fig:MassBH}
\end{figure*}

\subsection{The Supermassive Black Hole in GHZ9}\label{sec:SMBH}
The X-ray detection provides evidence that GHZ9 hosts an accreting SMBH. 
We derive a bolometric luminosity of 2.0 (0.6) $\times 10^{46}$\ergse for $\Gamma$ = 2.3 ($\Gamma$ = 1.8), assuming standard bolometric corrections for quasars \citep{duras20}. This bolometric luminosity is consistent with an alternative estimate derived from the continuum at 4400 \AA\ \citep{duras20}, after correcting for the measured extinction value (see Section~\ref{sec:physical}), thus pointing to a significant AGN contribution to the optical emission.

The bolometric luminosity derived implies that, if the BH in GHZ9 is radiating at its Eddington limit (f$_{Edd}$ = 1), its mass is  1.60$\pm$0.31 (0.48$\pm$0.09) $\times 10^8\msun$, consistent with the value reported by \cite{Kovacs2024}. We note that the reported uncertainty in the BH mass only includes the 0.5--7 keV error counts. Additional uncertainties, due to the intrinsic scatter in the bolometric corrections, could be as high as a factor of $\sim$ 2 \citep[see Figure 2 in][]{duras20}. \\
Accretion at lower Eddington rates would imply higher BH masses, even when accounting for smaller bolometric corrections \citep{lusso12}.
Instead, if GHZ9 is a super-Eddington accretor with funnel-like geometry \citep{king24} and is seen along the funnel (face-on), the same observed luminosity might be produced by a BH that is 10--100 times smaller. 
However, it has also been suggested that at high accretion rates, the hot X-ray emitting plasma undergoes a large photon supply from the accretion disk and the funnel walls. This would Compton-cool the plasma down to $\approx 10$ times lower values than in standard AGN coronae, resulting in reduced X-ray emission \citep{madau24}, consistent with the widespread X-ray weakness of high-redshift AGN discovered by JWST \citep{ananna24, Mazzolari2024b, maiolino24}. In this scenario, the X-ray bolometric corrections would be $\approx 10$ times larger than in standard quasars, leading to a bolometric luminosity for GHZ9 that is so high that it would again require a BH mass of $\approx 10^8\msun$ to power it. 

\subsection{Accretion History and BH-mass-to-stellar-mass Ratio}
If Eddington-limited, the BH mass estimated for GHZ9 would require an initial seed of $\sim$10$^6$ M$_{\odot}$ at z=18 \citep[yellow solid line from Figure~\ref{fig:MassBH}, see][]{Valiante2016}. 
Allowing for past super-Eddington accretion with f$_{Edd}$ = 1.5 (f$_{Edd}$ = 2) would alleviate the need to grow the BH in GHZ9 from a 10$^6$ M$_{\odot}$ heavy seed already in place at z=18, reducing the required masses to lighter seeds of 10$^4$ M$_{\odot}$ (10$^2$ M$_{\odot}$) at z = 25 \citep{Valiante2016}, as shown by the pink dashed (blue dot-dashed) line in Figure~\ref{fig:MassBH}.
We note that \cite{Dayal2024} and \cite{Huang2024} proposed a sub-Eddington accretion scenario onto a supermassive primordial BH seed of $\sim$10$^4$ M$_{\odot}$ for GHZ9.

The high BH mass points towards a high M$_{BH}$/M$_{star}$ ratio (Figure~\ref{fig:MassBH}, right panel). When considering M$_{star}$ obtained with the best-fit f$_{AGN}$=0.3, the M$_{BH}$ derived for $\Gamma$ = 2.3 ($\Gamma$ = 1.8) implies an M$_{BH}$/M$_{star}$= 0.33$\pm$0.22 (0.10$\pm$0.07). 
Such a high ratio is in agreement with the analysis in \citet{Kovacs2024}, and consistent with typical values measured in high-redshift AGN \citep[e.g.,][]{Furtak2024, Maiolino2023}.

We note that the M$_{BH}$/M$_{star}$ ratio is affected by significant systematic uncertainties due to the assumptions made in deriving M$_{BH}$, and to the wide range of M$_{star}$ values associated with the observed photometry at varying AGN contributions (Section~\ref{sec:physical}). 
We first aim at deriving a stringent lower limit on M$_{BH}$/M$_{star}$. The maximal estimate of the stellar mass is obtained in the case f$_{AGN}$=0 (M$_{star}$ = 7.2$^{+3.0}_{-3.8}$ $\times$ 10$^8$ M$_{\odot}$). We stress that this scenario is conservative because both the SED fitting and the 4400~\AA\ continuum suggest a non-negligible AGN contribution to the SED. We assessed that this value is robust against additional systematics in the assumed IMF and SFH. In fact, the adoption of a top-heavy IMF, which is considered to be more appropriate for a low-metallicity object as GHZ9 \citep[e.g.,][]{Chon2022,Trinca2024}, would decrease the stellar mass estimate. Unsurprisingly, the SFH is poorly constrained, but the fit obtained for GHZ9 does not appear to be significantly affected by the \textquote{outshining} effect, which may lead to an underestimate of M$_{star}$ \citep{GimenezArteaga2024}. Indeed, it predicts $>$50\% of the mass to have formed in a 400 Myr old burst, and the uncertainty considers the case with a fraction as low as 10\% of M$_{star}$ forming in the ongoing burst. 
We then consider the most conservative M$_{BH}$ estimate, which is obtained under the assumptions of unabsorbed emission with f$_{Edd}$=1, $\Gamma$ = 1.8 and including both the nominal uncertainty on the X-ray flux and a factor of 2 uncertainty on the bolometric correction (M$_{BH}$ = 4.8$^{+4.8}_{-2.2}$ $\times$ 10$^7$ M$_{\odot}$). We find the stringent lower limit on M$_{BH}$/M$_{star}$ is 0.07$^{+0.1}_{-0.05}$. 
The lower bound M$_{BH}$/M$_{star}>$2\% (red star in Figure~\ref{fig:MassBH}) is significantly higher than expected from the \citet{Kormendy2013} relation, and in line with the accelerated growth of BHs relative to stellar mass observed in high-redshift AGN. 
Much higher ratios are obtained when relaxing the aforementioned assumptions, i.e., a steep spectral index $\Gamma$ = 2.3 leads to an M$_{BH}$/M$_{star}$ ratio consistently $>$0.22, and as high as $\simeq$ 0.48 when assuming the SED to be 60\% (f$_{AGN}$=0.6) contributed by AGN emission. 

\section{Conclusions} \label{sec:Summary}
The combined analysis of NIRSpec and Chandra data clearly indicates that GHZ9 hosts an AGN at z=10.145. A robust association to a statistically significant point-like X-ray emission in Chandra data demonstrates the presence of AGN emission originating from an accreting SMBH. Under standard assumptions, the X-ray luminosity corresponds to a BH mass of $\sim$0.5-1.6 $\times$ 10$^8$ M$_{\odot}$. We caution that additional uncertainties as high as a factor of $\sim$2 could arise from the intrinsic scatter in the bolometric corrections adopted. The new JWST NIRSpec PRISM results that we present here are used to inspect a number of emission line diagnostics. From the resulting UV line ratios diagnostics, the dominant ionizing source could be either AGN or star formation, while EW diagnostics suggest that an AGN is the more likely scenario. We find that GHZ9 is metal--poor (Z$<$0.1 Z$_\odot$), and significantly N-enhanced (6--9.5 (N/O)$_{\odot}$), while its C/O is sub-solar (0.2--0.65 (C/O)$_{\odot}$). The measured spatial extension in NIRCam images shows that the object is also very compact (r$_e<$ 106 pc). 

The inferences obtained from the X-ray and SED analysis suggest intriguing scenarios for the formation of SMBH and the coevolution with their host galaxies. Our stringent limit of  M$_{BH}$/M$_{star}>$0.02 indicates an accelerated evolution of the BH mass compared to the stellar mass. However, without constraints on the Eddington ratio, the scenarios regarding the initial BH seed remain uncertain, as in the case for other high-redshift AGN \citep{Larson2023B,Bogdan2024,Maiolino2023}.

We also note that GHZ9 belongs to a well-defined, homogeneous photometric sample selected based on pure Lyman-Break color selection criteria \citep{Castellano2022b,Castellano2023}. As described in \citet{Napolitano2024B}, program GO-3073 confirmed all the six objects from the parent GLASS-JWST sample at $z\geq 9.5$, plus two additional sources from alternative photometric selections. Out of these objects, at least GHZ9 is a confirmed AGN. Despite the low--number statistics, this is in line with the growing consensus that about 10-15\% of bright, high-redshift galaxies host AGNs \citep[e.g.][]{Maiolino2024}.

It is also tempting to place these findings in the context of other high-redshift galaxy observations. The most obvious analog is GHZ2, another bright object at $z\simeq 12.3$, which displays strong UV emission lines \citep{Castellano2024}. As shown in Figure~\ref{fig:AGNvsSF_diagrams}, GHZ2 and GHZ9 are essentially indistinguishable in terms of UV line ratios and EW - the only notable difference being the He~II line, with an EW$\simeq$18~\AA\ in GHZ9 which is significantly higher than in GHZ2 (EW$\simeq$5~\AA). In both cases, these objects are compact and N--enhanced, with a C/O ratio in line with expectations for low-metallicity galaxies at high redshift.
Unfortunately, lacking an X-ray analysis and high-resolution JWST spectra, it is not possible to ascertain whether GHZ2 also hosts an AGN - future observations are clearly needed to investigate this.

GHZ9 also shares common properties with the bright object GN-z11 at z=10.6, which is compact and N-enhanced, albeit with a less extreme UV spectrum \citep{Bunker2023B}, and has broad-line components and extreme gas densities, indicating the presence of a BLAGN \citep{Maiolino2023}.

Based on their UV line-emission properties alone, GHZ9 (and GHZ2) are similar to the class of strong C~IV emitters at lower redshifts \citep[][]{Izotov2024,Topping2024,Topping2024b}. There is no evidence that these objects host an AGN, and their line ratios seem to indicate that they experience a dense starburst.
This leaves open the possibility that many, if not all, of these newly-discovered sources also host AGN, but their active nuclei remain elusive due to X-ray weakness, beamed, or absorbed emission \citep[e.g.,][]{Madau2024,Maiolino2024}.
A comprehensive scenario connecting GHZ9, other strong C~IV emitters such as GHZ2, and the broader category of N-enhanced objects, including GN-z11 and others \citep{Isobe2023,Schaerer2024,Topping2024b} is currently missing. However, a composite nature involving both AGN activity and star formation in a low-metallicity, dense interstellar medium (ISM) enriched by massive or supermassive stars appears consistent with the available evidence \citep{Charbonnel2023,DAntona2023,Boorman2024,Calabro2024B,MarquesChaves2024,Zavala2024}.

These scenarios deserve an in-depth analysis with future observations. In this context, GHZ9, which  is the only z$>$9 object showing both a highly-ionizing, N-enhanced spectrum and X-ray emission, is the ideal target to improve constraints on SMBH-host coevolution and SMBH seeding mechanisms.
Specifically, we can individuate two main directions for progress:
(1) constraining the physical properties of the host, and the contribution of the AGN to the total UV/optical emission; and (2) obtaining independent constraints on M$_{BH}$, and thus the Eddington ratio. 


The physical conditions of the star-forming ISM (i.e., density, temperature, abundances) can be assessed through high-resolution spectroscopy, to accurately constrain the origin of the copious amounts of ionizing photons \citep[e.g.,][]{Ji2024}. Medium or high-resolution NIRSpec spectroscopy, as well as MIRI Medium Resolution Spectroscopy, can search for broad-line components in the permitted lines of GHZ9 and other similar sources to assess the AGN fraction, currently the main uncertainty in deriving M$_{BH}$/M$_{star}$. Most importantly, estimating M$_{BH}$ from broad lines would allow determination of the Eddington ratio in combination with the X-ray luminosity \citep[e.g.,][]{Lusso2010}, providing a direct constraint on the seeding mechanisms of the SMBH in GHZ9. Finally, ALMA can measure the  dynamical mass using FIR lines, enabling direct comparison between M$_{BH}$/M$_{star}$ and M$_{BH}$/M$_{dyn}$, which would constrain the relative timescales of BH and stellar-mass growth \citep[e.g.,][]{Pensabene2020}. 


Significant progress in determining the X-ray properties of GHZ9 and similar objects \citep[e.g., UHZ1][]{Goulding2023} will have to await next-generation X-ray imaging satellites with large collecting areas and $\sim$arcsec angular resolution, such as the Advanced X-ray Imaging Satellite, a probe-class mission currently under evaluation at NASA \citep{marchesi20,reynolds23}. For the time being, as in the case of GHZ9, a detailed analysis of JWST-selected candidates discovered in fields with deep Chandra imaging will be essential to further constrain different scenarios on the early stages of galaxy-AGN coevolution.

\begin{acknowledgments}
We thank the referee for the constructive feedback provided that helped us improve this paper. 
We thank J. S. Dunlop, R. Ellis, E. Giallongo, A. Trinca, and R. Valiante for the interesting discussions. We are grateful to K. Nakajima and A. Feltre for kindly providing updated tables of their line-emission models.
This work is based on observations made with the NASA/ESA/CSA James Webb Space Telescope (JWST). The JWST data presented in this article were obtained from the Mikulski Archive for Space Telescopes (MAST) at the Space Telescope Science Institute. The specific observations analyzed are associated with program JWST-GO-3073 and can be accessed via \dataset[doi: 10.17909/4r6b-bx96]{https://doi.org/10.17909/4r6b-bx96}. We thank Tony Roman (program coordinator) and Glenn Wahlgren (NIRSpec reviewer) for the assistance in the preparation of GO-3073 observations. This paper employs a list of Chandra data sets, obtained by the Chandra X-ray Observatory, contained in~\dataset[doi: 10.25574/cdc.386]{https://doi.org/10.25574/cdc.386}. We acknowledge support from INAF Mini-grant 2022 "Reionization and Fundamental Cosmology with High-Redshift Galaxies", INAF Mini-grant 2022 “The evolution of passive galaxies through cosmic time”, INAF Large Grant 2022 “Extragalactic Surveys with JWST” (PI: Pentericci), INAF Large GO grant "AGNpro" (PI: Vito), INAF Large GO 2023 grant "AGNpro" (PI: Vito), PRIN 2022 MUR project 2022CB3PJ3 - First Light And Galaxy aSsembly (FLAGS) funded by the European Union – Next Generation EU and the European Union – NextGenerationEU RFF M4C2 1.1 PRIN 2022 project 2022ZSL4BL INSIGHT. L.Z. and E.P. acknowledge support from the Bando Ricerca Fondamentale INAF 2022 Large Grant “Toward an holistic view of the Titans: multi-band observations of z$>$6 QSOs powered by greedy supermassive black holes". L.Z. acknowledges support from
the European Union – Next Generation EU, PRIN/MUR 2022
2022TKPB2P – BIG-z. T.T. and S.R.R. acknowledge support from NASA through grant JWST-GO-3073. K.G. and T.N. acknowledge support from Australian Research Council Laureate Fellowship FL180100060. Support was also provided by NASA through grant JWST-ERS-1342.
\end{acknowledgments}

\facilities{JWST}


\software{Astropy \citep{Astropy2013}, Matplotlib \citep{Hunter2007}, Specutils (\url{https://specutils.readthedocs.io/en/stable/}), \textsc{emcee} \citep{Foreman_Mackey2013}, \textsc{CIGALE} \citep{Boquien2019}, \textsc{Galight} \citep{Ding2020}}



%
    


\bibliography{biblio}{}
\bibliographystyle{aasjournal}



\end{document}